\documentclass[a4paper,11pt]{article}

\usepackage{amsmath,amssymb,color,graphics,epsfig,cite}
\usepackage{titlesec}
\usepackage[titletoc,title]{appendix}
\usepackage{hyperref}
\usepackage{xcolor}  % Required for defining and using colors
\usepackage{soul}    % For highlighting text with \hl (handles line breaks)
\usepackage{empheq}  % For highlighting entire equations (requires amsmath)
\sethlcolor{red} 
\soulregister{\cite}{7} % 注册\cite命令
\soulregister{\citep}{7} % 注册\citep命令
\soulregister{\citet}{7} % 注册\citet命令
\soulregister{\ref}{7} % 注册\ref命令
\soulregister{\pageref}{7} % 注册\pageref命令

\usepackage{amsfonts}
\newcommand{\hoch}[1]{$\, ^{#1}$}

\newcommand{\be}{\begin{equation}}
	\newcommand{\ee}{\end{equation}}
\newcommand{\bea}{\setlength\arraycolsep{2pt} \begin{eqnarray}}
	\newcommand{\eea}{\end{eqnarray}}
\newcommand{\nn}{\nonumber}

\def\ft#1#2{{\textstyle{\frac{\scriptstyle #1}{\scriptstyle #2} } }}
\def\fft#1#2{{\frac{#1}{#2}}}

\def\0{{\sst{(0)}}}
\def\1{{\sst{(1)}}}
\def\2{{\sst{(2)}}}
\def\3{{\sst{(3)}}}
\def\4{{\sst{(4)}}}
\def\5{{\sst{(5)}}}
\def\6{{\sst{(6)}}}
\def\7{{\sst{(7)}}}
\def\8{{\sst{(8)}}}
\def\sst#1{{\scriptscriptstyle #1}}

\def\a{\alpha}

\def\e{\epsilon}
\def\l{\lambda}
\def\d{\delta}
\def\p{\phi}

\def\pd{\partial}
\def\O{\mathcal{O}}

\def\ext{\text{ext}}
\def\At{\tilde{A}}
\begin{document}
	
	\begin{center}
		{\Large {\bf Weak Cosmic Censorship Conjecture in Gedanken Experiments at All Orders}}
		
		\vspace{20pt}

Kai-Peng Lu\hoch{1}, Peng-Yu Wu\hoch{1} and H. L\"{u}\hoch{1,2}
		
		\vspace{10pt}
		
		{\it \hoch{1}Center for Joint Quantum Studies, Department of Physics,\\
			School of Science, Tianjin University, Tianjin 300350, China }
		
		\bigskip
		
{\it \hoch{2}The International Joint Institute of Tianjin University, Fuzhou,\\ Tianjin University, Tianjin 300350, China}
		
		\vspace{40pt}
		
\underline{ABSTRACT}
	\end{center}

We present a systematic analysis of the Weak Cosmic Censorship Conjecture (WCCC) through Gedankenexperiments involving black hole perturbations induced by test particles. Our approach allows for the calculation of perturbations to any order for a general class of black holes that admit the zero-temperature extremal limit. We find that the WCCC for extremal and near-extremal black holes is hinged upon the positive sign of only one quantity, namely $W=\left( \frac{\pd S}{\pd T} \right)_{Q_\alpha ; T=0}$, which is indeed positive for all the well-known black holes.

\vfill {\footnotesize kaipenglu@tju.edu.cn \ \ wupy2023@tju.edu.cn\ \ \ mrhonglu@gmail.com}

\thispagestyle{empty}
\pagebreak

%\tableofcontents
%\addtocontents{toc}{\protect\setcounter{tocdepth}{2}}

\newpage

\section{Introduction}

The Weak Cosmic Censorship Conjecture (WCCC), proposed by Roger Penrose in 1969 in the context of Einstein's General Relativity, posits that naked singularities cannot form in realistic gravitational systems\cite{Penrose:1969pc}. Instead, any singularity resulting from gravitational collapse must be hidden within an event horizon, therefore preserving the predictability of the laws of physics\cite{Singh:1997wa}. While it remains unproven, WCCC has significant implications for the study of black holes and General Relativity.

One approach to examining potential violations of WCCC is through perturbation theory to determine whether naked singularities can form. Wald's pioneering work considered a gedankenexperiment \cite{Wald:1974hkz} involving a test particle with electric charge and angular momentum falling into extremal Kerr-Newman (KN) black holes. Wald demonstrated that the extremal KN black holes do not absorb the particle that would cause naked singularity due to electromagnetic and centrifugal repulsion forces, ensuring the validity of WCCC. However, Hubeny \cite{Hubeny:1998ga} later proposed that WCCC could be violated for slightly non-extremal Reissner-Nordstr\"om (RN) black holes (Hubeny’s type).  The same violation was repeated for the nearly extremal KN black holes as well \cite{Jacobson:2009kt}. Further developments have shown \cite{Gao:2012ca} that WCCC could also be apparently violated even for the extremal KN black holes by expanding the horizon condition up to the second-order terms of mass $\d M $ and charges $\d Q_\a $ (mixed type), but without including the $\delta^2 M$ and $\delta^2 Q_\a $ contributions. (Here $ Q_\alpha $ denotes all conserved charges.) Refer to \cite{HT1,HT2,HT3,HT4,HT5} for further studies on Hubeny's type, and consult \cite{HT4,HT5,MT1,MT2,MT3,MT4,MT5,MT6} for works on the mixed type. (See also \cite{Ghosh:2019dzq,Ghosh:2021cub} for the study of WCCC in modified gravities.)

To reconcile these conflicting results, Sorce and Wald (SW) \cite{Sorce:2017dst,Wald:2018xxi} showed that the linear-order analysis in Hubeny's gedankenexperiment \cite{Hubeny:1998ga} was insufficient, as the potential violation in the near-extremal expansion must be of the second order \cite{Shaymatov:2019del,Shaymatov:2020byu}. With the full second-order terms included, there is no violation of the WCCC for the near-extremal KN black holes at this second order \cite{Sorce:2017dst,Wald:2018xxi}.

These case-by-case studies of WCCC in Gedankenexperiments were extended \cite{Wu:2024ucf} to a general class of black holes that are parameterized independently by mass and a set of charges $Q_\alpha$. These black holes are also assumed to have a zero-temperature extremal limit, where the mass becomes a function of the charges. It was shown that, up to and including the second-order perturbation, the WCCC is hinged upon the sign of the quantity \cite{Wu:2024ucf}
\begin{equation} \label{W1}
	W \equiv \left( \frac{\pd S}{\pd T} \right)_{Q_\alpha;T=0}.
\end{equation}
Specifically, $W > 0$ preserves WCCC, whereas $ W < 0$ suggests a potential violation. Intriguingly, all the well-known black holes, including the RN and KN black holes, have positive $W$. It was shown in \cite{Wu:2024ucf} that black holes with negative $W$ could only arise in black holes that violate the no-hair theorem. It was further argued that spherically-symmetric and static black holes necessarily have positive $W$.

In this paper, we would like to extend the discussion to the higher-order perturbations, by refining the procedure introduced in \cite{Wu:2024ucf}. This is a necessary investigation since there can exist parameters so that the second-order perturbation vanishes, in which case the higher-order terms become important. In these cases, would $W$ still be the determining factor to protect WCCC, or will some new quantities enter the game? It turns out that higher-order perturbation in Gedankenexperiments for the KN black hole was investigated by Wang and Jiang (WJ) \cite{Wang:2022umx}. This provides a useful check of our general approach. However, although our general formulae confirm that the WJ's conclusion for the KN black hole is morally correct, we disagree with their results in some specific detail.

The paper is organized as follows. In Sec.~\ref{sec:setup}, we set up the investigation by properly defining the horizon condition for the WCCC. A key modification compared to the earlier work \cite{Wu:2024ucf} is that we insist on using entropy and charges as thermodynamic variables in the canonical ensemble where the mass is the thermodynamic potential. We also define a physical process as one that follows the second law, namely that the total change of the entropy, including perturbative contributions of all orders, must be nonnegative. In Sec.~\ref{sec:orders}, we investigate the perturbative Gedankenexperiments order by order. We begin with a review of the first and second order, but in our new language. We modify the technique of \cite{Wu:2024ucf} so that we can proceed to investigate the perturbations to all higher orders. In Sec.~\ref{sec:direct}, inspired by the earlier results, we use a simpler one-step calculation and obtain the general results that encompass all the previous order-by-order discussions. In Sec.~\ref{sec:KN}, we revisit the KN black hole and compare our results to those of \cite{Wang:2022umx}. We conclude the paper in Sec.~\ref{sec:con}. In appendix \ref{app:notation}, we summarize the key notation and conventions used in this paper.

\section{The Setup} \label{sec:setup}

The original WCCC in Gedankenexperiment proposed by Wald showed that the extremal Kerr-Newman black hole was safe. The apparent violation in the near-extremal region \cite{Hubeny:1998ga} was later resolved in \cite{Sorce:2017dst,Wald:2018xxi}, by considering the second-order perturbation which becomes nonnegligible in the near but yet extremal region. This SW procedure has been generalized \cite{Wu:2024ucf} to a general class of black holes that have zero-temperature extremal limit and are parameterized by the mass and a set of charges. The WCCC in Gedankenexperiment at the second order was shown to be predicated upon the positive sign of $W$. In this paper we shall extend the work of \cite{Wu:2024ucf} to all orders.

To organize the perturbative calculation, we follow \cite{Sorce:2017dst,Wald:2018xxi,Wu:2024ucf} and introduce an order parameter $\lambda$. For example, the mass $M$ of a black hole after the perturbation $\Delta M$ can be written as
\be
M + \Delta M = M(\lambda) = M + \lambda \delta M + \ft12 \lambda^2 \delta^2 M + \cdots + \fft1{n!} \lambda^n \delta^n M+ \cdots .
\ee
Note that the parameter $\lambda$ is used solely for the bookkeeping purpose, where the power $n$ in $\lambda^n \delta^n$ signifies the $n^{\rm th}$-order contribution to the expansion. It is understood that $\lambda$ will be eventually set to 1. Therefore $\delta^n M$ above denotes the $n^{\rm th}$-order perturbation of the mass, and the total mass perturbation is
\be
\Delta M= \lambda \delta M + \ft12 \lambda^2 \delta^2 M + \cdots + \fft1{n!} \lambda^n \delta^n M+ \cdots\,.\label{Deltadef}
\ee
The same formalism applies to all other thermodynamic quantities of the black hole. It is worth pointing out that if there exists a full non-perturbative close expression $M(\lambda)$, we have $\delta^n M = \partial^nM/\partial\lambda^n$ evaluated at $\lambda=0$. However, such an explicit expression is unlikely to exist and it is also unnecessary in our perturbative approach. However, we make an {\it a priori} assumption here that $M(\lambda)$ is analytic so that our discussion of the perturbation can be extended to arbitrarily higher orders. For a specific black hole, such an assumption may not be valid, and we shall come back to this point later to discuss what black hole properties determine how high the order we can pursue in this perturbative approach.

The thermodynamic quantities of a black hole are not all independent. Wald's analysis of the Kerr-Newman black hole treats the temperature and charges $(Q(\lambda),J(\lambda))$ as independent variables, and furthermore, the near-extremal temperature $T_\epsilon$ is taken to be the same order as $\lambda$. In other words, a dimensionless parameter $\epsilon\sim \lambda$ is introduced so that $T_\epsilon\propto \epsilon$. This strategy works well for the Kerr-Newman black hole, where thermodynamic quantities can be expressed analytically in terms of each other. However, for a general class of black holes, we do not have such explicit relations, except for the thermodynamic first and second laws. In a canonical ensemble, the mass is a function of the entropy $S_\epsilon$, not the temperature $T_\epsilon$, together with a set of charges $Q_\alpha$. Although it is equivalent to treat either $S_\epsilon$ or $T_\epsilon$ as thermodynamic variables, the derivation becomes less convoluted in the former case for the canonical ensemble.

With these preliminaries, we can now state the horizon condition for the WCCC. We consider a black hole of mass $M$ and charges $Q_\alpha$, where $\alpha=1,2,\cdots, N$, labels the $N$ number of charges. In the near-extremal region, the mass and charges are related by $M=M(S_\epsilon, Q_\alpha)$. In the extremal limit, both the mass and entropy become functions of charges $Q_\alpha$ solely:
\be
M_{\rm ext}(Q_\alpha) = M(S_{\rm ext}(Q_\alpha), Q_\alpha)\,.
\ee
We assume that after a certain physical process, the mass and charges of the spacetime geometry become $M+\Delta M$ and $Q_\alpha + \Delta Q_\alpha$. We can define a quantity $X_\epsilon$, as in \cite{Wu:2024ucf}:
\be
X_\epsilon \equiv M(S_\epsilon, Q_\alpha) + \Delta M - M_{\rm ext} (Q_\alpha + \Delta Q_\alpha)\,.\label{horizoncond}
\ee
The WCCC is then protected if $X_\epsilon\ge 0$. Here we have assumed that naked singularity only arises if the mass of the spacetime is less than the extremal black hole mass for given charges $Q_\alpha$. Implicitly we also assume that the black hole continues to be if the mass is bigger, although this is not essential, since we would not rule out the possibility that horizonless regular geometries might also arise, which would not affect our derivation and conclusion. Nevertheless, we shall refer to $X_\epsilon \ge 0$ as the horizon condition. It follows from the discussions above, we have
\be\label{Xepsilon}
X_\epsilon = M\big(S_\epsilon(\lambda), Q_\alpha(\lambda)\big) -
M\big(S_{\rm ext}(Q_\alpha(\lambda)), Q_\alpha(\lambda)\big)
=M(\lambda) - M\big(S_{\rm ext}(Q_\alpha(\lambda)), Q_\alpha(\lambda)\big)\,.
\ee
In order to evaluate $X_\epsilon$, it is necessary to have a meaningful definition of a ``physical process.'' For the RN and KN black holes, there are three approaches that lead to the same conclusion. One is based on the future-pointing ($\dot t\ge 0$) geodesic motion \cite{Wald:1974hkz, Gao:2012ca,MT4}. The other is based on the null-energy condition (NEC) of the matter crossing the horizon \cite{HT1}. (Also see the earlier work of \cite{poisson}.) Both lead to the second law of the black hole thermodynamics, $\delta S\ge 0$. The first and the second approaches to defining the physical process are strongly model-dependent and cannot apply to the discussion of the general class of black holes. In this paper, we shall apply the (model-independent) second law, namely
\be
\Delta S\ge 0\,,\label{secondlaw}
\ee
where $\Delta S$ is defined the same as \eqref{Deltadef}, but with $M$ replaced by $S$.  In other words, the total entropy change, including all orders contributions, should be nonnegative under any physical process. The order-by-order approach is as follows. The leading order is simply $\delta S\ge 0$. When it is saturated, we need to consider the second order and require $\delta^2 S\ge 0$. Generally speaking, the $n^{\rm th}$ order becomes relevant when the first $(n-1)^{\rm th}$ orders are saturated, {\it i.e.}
\be
\delta S = \delta ^2 S = \cdots = \delta^{n-1} S =0\,,\qquad
\delta^n S\ge 0\,.\label{nordercond}
\ee
The case of the second-order perturbation $(n=2)$ was first proposed in \cite{Lin:2024deg}. In the canonical ensemble, the mass $M$ is the function of entropy $S$ and charges $Q_\alpha$. We now examine this relation perturbatively order by order. With the order parameter $\lambda$, we have
\be
M(\lambda) - M(S(\lambda), Q_\alpha(\lambda))=0\,.\label{MSQ}
\ee
The leading order of $\lambda$ is simply the identity. The first order leads to the first law of thermodynamics
\be
\delta M - \phi_\alpha \delta Q_\alpha = T \delta S + {\cal O}(\lambda^2)\,,
\ee
where $\phi_\alpha = (\partial M/\partial Q_\alpha)_S$ and $T = (\partial M/\partial S)_Q$. Note that in this paper, when a Greek letter labelling charges appears twice in a product, summation over all charges is assumed. The requirement of $\delta S\ge 0$ implies that
\be
\delta M - \phi_\alpha \delta Q_\alpha\ge 0\,.
\ee
For the second order, we impose $\delta S=0$ and $\delta ^2 S \ge 0$, as in \cite{Lin:2024deg}, we have
\be\label{deltaM}
\delta M - \phi_\alpha \delta Q_\alpha=0\,,\qquad
\delta^2 M - \phi_\alpha \delta^2 Q_\alpha\ge \delta\phi_\alpha \delta Q_\alpha\equiv Y^\2\,.
\ee
It was shown \cite{Wu:2024ucf} that the quantity $Y^\2$ is simply the Hessian metric of $M(S, Q_\alpha)$, denoted as $DM$. It conforms to the Hessian metric of $S(M, Q_\alpha)$:
\be
DM= - T DS = -\fft{\kappa}{8\pi}\, D{\cal A}\,,
\ee
where $\kappa$ and ${\cal A}$ are the surface gravity and area of the horizon respectively. In \cite{Sorce:2017dst}, where KN black hole was analyzed, the Hessian metric $D{\cal A}$ is denoted as $\delta^2 {\cal A}$. Thus, in the language of \cite{Sorce:2017dst}, the second-order inequality is expressed as
\be
\delta^2 M - \phi_\alpha \delta^2 Q_\alpha\ge -\fft{\kappa}{8\pi}\, \delta^2{\cal A}\,.
\ee
(Since this $\delta^2$ appearing in front of ${\cal A}$, and its higher-order generalization, does not fit the meaning of $\delta^n$ defined in \eqref{Deltadef}, we shall not generally use this notation of \cite{Sorce:2017dst}.) For the general $n^{\rm th} $-order condition \eqref{nordercond}, we have
\bea
&&\delta^i M - \phi_\alpha \delta^i Q_\alpha = Y^{\sst{(i)}}\,,\qquad i=1,2,\ldots, n-1\,,\nn\\
&&\qquad\qquad\delta^n M - \phi_\alpha \delta^n Q_\alpha \ge  Y^{\sst{(n)}}\,.
\label{physcondn}
\eea
where
\be
Y^\1=0\,,\qquad Y^{\sst{(k)}} =\sum^{k-1}_{i = 1} C^i_{k-1} \d^i \phi_\alpha \d^{k-i} Q_\alpha\,,\qquad k\ge 2\,.
\ee
Here, $C^i_j $ with $j\ge i$ is the binomial coefficient.

With these preliminaries, we are in the position to evaluate the WCCC Gedankenexperiments order by order. Specifically, we shall apply the condition \eqref{physcondn} to determine the sign of $X_\epsilon$ defined in \eqref{Xepsilon}. We shall carry it out in the next section.

\section{Gedankenexperiments order by order}
\label{sec:orders}

In this section, we explicitly evaluate the $X_\epsilon$ under the condition \eqref{physcondn}.  It follows from \eqref{Xepsilon} that up to and including the $n^{\rm th}$ order, we have
\begin{equation}
	X_\epsilon = M(S_\epsilon, Q_\alpha) - M_\ext (S_\ext, Q_\alpha) + \sum_{i=1}^{n} \frac{\lambda^i}{i!} \left( \delta^i M - \delta^i M_\ext \right) + {\cal O}(\lambda^{n+1})\,, \label{X-gen}
\end{equation}
where $\delta^i M$'s with $i=1,2,\ldots, n$ are subject to the conditions given in \eqref{physcondn}. There is an additional near-extremal expansion in terms of temperature $T_\epsilon$. Thus we have
\begin{equation}
S_\e = S_\ext + \sum_{i=1}^{\infty} \frac{1}{i!} W_i\, T_\e^i\,,\qquad
W_i \equiv \left( \frac{\pd^i S}{\pd T^i} \right)_{Q_\alpha;T=0}\,.\label{Wn}
\end{equation}
Note that $W_1=W$ as defined in \eqref{W1}. Thus while $M_\ext$ and $\delta^i M_{\rm ext}$ do not involve $T_\epsilon$, $M$ and $\delta^i M$, which are functions of $S_\epsilon$, all have additional power expansions involving $T_\epsilon$. For simplicity, we follow Sorce and Wald \cite{Sorce:2017dst,Wald:2018xxi} and consider homogeneous polynomials of $\lambda$ and $T_\epsilon$. In other words, we treat the temperature $T_\epsilon$ as having the same order as $\lambda$; otherwise, the analysis could be prone to errors that could lead \cite{Wu:2024ucf} to incorrect results corresponding to the Hubeny's type and mixed type. However, it should be stressed that they do not have to be in the same order, and in Sec.~\ref{sec:direct}, we generalize the assumption and obtain the WCCC condition for which $\lambda$ and $T_\epsilon$ may not have to be homogeneous. The general expansion is complicated and we shall analyze the perturbation starting from simpler lower orders to the more complicated higher ones.

We now come back to the question raised earlier: how high the perturbative order we can pursue in this WCCC analysis. In the series expansion \eqref{Wn}, we have assumed that entropy is infinitely differentiable in terms of the temperature in the extremal $T_\epsilon \rightarrow 0$ region. This is indeed true for the KN black hole even though its $\partial S/\partial T$ can be divergent at some specific nonzero temperature. In general if the function $S(Q_\alpha, T_\epsilon)$ is only $k^{\rm th}$ differentiable at $T_\epsilon=0$, then we can only discuss the WCCC perturbation up to the $n\le k$ order. It should be pointed out however that non-infinitely differentiable black holes in the extremal limit are rare in literature, if they exist at all.

\subsection{The first and second-order perturbations}

At the first order $(n=1)$, Eq.~\eqref{X-gen} under \eqref{physcondn} simply yields
\be
X_\epsilon \ge (\phi_\alpha - \phi^{\ext}_\alpha)\, \delta Q_\alpha + {\cal O}(\lambda^2) = 0 + {\cal O}(\lambda^2)\,.
\ee
This is because the quantity in the bracket must be at order $T_\epsilon$ and hence the first term is beyond the first order. This is of course the result of \cite{Wald:1974hkz} for the extremal KN black hole. The involvement of the temperature in the near-extremal region is necessarily at least second order in the expansion.

At the second order $(n=2)$, Eq.~\eqref{X-gen} under \eqref{physcondn} becomes
\bea
	X_\e &=& M(S_\e, Q_\alpha) - M_\ext (S_\ext, Q_\alpha) + \l (\d M - \d M_\ext) + \ft12\l^2 (\d^2 M - \d^2 M_\ext) + \O(\l^3)\nn\\
&\ge& \Big(M(S_\e, Q_\alpha) - M_\ext (S_\ext, Q_\alpha)\Big) + \l \left( \p_\alpha - \p^\ext_\alpha \right) \d Q_\alpha \nn\\
		&&+ \ft12\l^2 \left[ \left( \d \p_\alpha - \d \p^\ext_\alpha \right) \d Q_\alpha + (\p_\alpha - \p^\ext_\alpha ) \d^2 Q_\alpha \right]+ \O(\l^3)\,.\label{WCCC20}
\eea
A careful evaluation leads to the inequality
\begin{equation} \label{WCCC2}
	X_\e \ge \ft12 W_1 \Big(T_\e - \l \fft{\d S_\ext}{W_1}\Big)^2 +\O(\lambda^3)\,,
\end{equation}
where $ \d S_\ext $ is defined as $ \d S_\ext = \frac{\pd S_\ext}{\pd Q_\alpha} \d Q_\alpha $ and $ W_1=W $. This is the result reported in \cite{Wu:2024ucf}, which is equivalent to the SW conclusion for the KN black hole.

\subsection{Generalizing to higher orders}

In the above derivation, we applied the condition \eqref{physcondn}, which is itself obtained by applying \eqref{nordercond} on the equality \eqref{MSQ}.  While the intermediate step \eqref{physcondn} is elegant, the derivation from \eqref{WCCC20} to \eqref{WCCC2} requires some nontrivial exercise involving differential identities \cite{Wu:2024ucf}. It can be tedious and cumbersome to use this approach for higher-order perturbations. It is actually advantageous, aided by a computer program, to evaluate $X_\epsilon$ in \eqref{Xepsilon} directly under the condition \eqref{nordercond}. For example, applying \eqref{nordercond} at the $n^{\rm th}$ order, the equality \eqref{MSQ} becomes
\bea
\delta M &=& \fft{\partial M}{\partial Q_\alpha} \delta Q_\alpha\,,\qquad
\delta^2M = \fft{\partial M}{\partial Q_\alpha} \delta^2 Q_\alpha +
\fft{\partial^2 M}{\partial Q_\alpha\partial Q_\beta} \delta Q_\alpha
\delta Q_\beta\,,\nn\\
\delta^3 M &=& \fft{\partial M}{\partial Q_\alpha} \delta^3 Q_\alpha +
3\fft{\partial^2 M}{\partial Q_\alpha\partial Q_\beta} \delta Q_\alpha
\delta^2 Q_\beta + \fft{\partial^3 M}{\partial Q_\alpha\partial Q_\beta\partial Q_\gamma} \delta Q_\alpha\delta Q_\beta \delta Q_\gamma\,,\nn\\
&\cdots&\nn\\
\delta^n M &\ge & \fft{\partial M}{\partial Q_\alpha} \delta^n Q_\alpha + \cdots\,.
\eea
For $n=2$, we have
\bea
X_\epsilon &\ge& \Big(\fft{\partial M}{\partial S}\Big)_{T=0}\, \Big((W_1 T_\epsilon -\lambda \delta S_\ext) + (W_2 T_\epsilon^2 - \ft12\lambda^2 \delta^2 S_\ext)\Big)\nn\\
&&+ \fft12\Big(\fft{\partial^2 M}{\partial S^2}\Big)_{T=0} \Big(W_1^2 T_\epsilon^2 - \lambda^2 (\delta S_\ext)^2\Big) + \lambda \Big(\fft{\partial^2M}{\partial S\partial Q_\alpha}\Big)_{T=0} \delta Q_\alpha (W_1 T_\epsilon - \lambda \delta S_\ext)\nn\\
&& + {\cal O}(\lambda^3)\,.\label{WCCC2new}
\eea
Note again that $T_\epsilon$ is of the same order as $\lambda$. Note also again that before taking the $T=0$ extremal limit, $M$ is a function of independent variables, entropy $S$ and charges $Q_\alpha$.  It is clear that we have
\bea
\Big(\fft{\partial M}{\partial S}\Big)_{T=0}&=&T=0\,,\qquad
\Big(\fft{\partial^2 M}{\partial S^2}\Big)_{T=0} = \Big(\fft{\partial T}{\partial S}\Big)_{T=0} \equiv U_1(Q_\alpha) = \fft{1}{W_1}\,,\nn\\
\Big(\fft{\partial^2M}{\partial S\partial Q_\alpha}\Big)_{T=0} &=&
\fft{d}{dQ_\alpha}\Big[ \Big(\fft{\partial M}{\partial S}\Big)_{T=0}\Big] -
\Big(\fft{\partial^2 M}{\partial S^2}\Big)_{T=0} \fft{\partial S_\ext}{\partial Q_\alpha}= -
\fft{1}{W_1} \fft{\partial S_\ext}{\partial Q_\alpha}\,.\label{WCCC2id}
\eea
Substituting the above identities into \eqref{WCCC2new}, we arrive precisely at the inequality \eqref{WCCC2}. This new approach no longer requires us to perform an intermediate step of converting $\partial M/\partial Q_\alpha$ to the chemical potentials, but rather lets us work on the thermodynamic potential $M(S,Q_\alpha)$ and its partial derivatives directly. This makes it easier to generalize to the higher orders.

\subsubsection{The third-order perturbation}

The proceeding to the third order is straightforward. For the condition \eqref{nordercond} with $n=3$, we have
\bea
X_\epsilon &\ge& \fft1{2W_1} (W_1 T_\e - \l \d S_\ext)^2 +
\fft{T_\epsilon}{2W_1} \Big(W_1 W_2 T_\epsilon^2 -\fft{\partial S_\ext}{\partial Q_\alpha}\big(\lambda W_2 T_\epsilon \delta Q_\alpha +\lambda^2 W_1 \delta^2 Q_\alpha\big)\Big)\nn\\
&&+ \ft16 T_\epsilon^3 W_1^3 \Big(\fft{\partial^3M}{\partial S^3}\Big)_{T=0} +
\ft12 W_1^2 T_\epsilon^2 \Big(\fft{\partial^3M}{\partial S^2 \partial Q_\alpha}\Big)_{T=0} \delta Q_\alpha\nn\\
&& +\ft12 W_1 T_\epsilon \Big(\fft{\partial^3 M}{\partial S\partial Q_\alpha\partial Q_\beta}\Big)_{T=0} \delta Q_\alpha \delta Q_\beta + {\cal O}(\lambda^4)\,.
\label{WCCC30}
\eea
Analogous to the set of identities \eqref{WCCC2id} used for the second-order perturbation, we have the following relevant identities in the evaluation \eqref{WCCC30}:
\bea
&&\Big(\fft{\partial^3M}{\partial S^3}\Big)_{T=0} =
\Big(\fft{\partial^2 T}{\partial S^2}\Big)_{T=0}\equiv U_2(Q_\alpha)\,,\nn\\
&&
\Big(\fft{\partial^3M}{\partial S^2 \partial Q_\alpha}\Big)_{T=0}=
\fft{d}{dQ_\alpha}\Big[\Big(\fft{\partial^2 M}{\partial S^2}\Big)_{T=0}\Big]-
 \Big(\fft{\partial^3M}{\partial S^3}\Big)_{T=0} \fft{\partial S_\ext}{\partial Q_\alpha}\,,\nn\\
&&
\Big(\fft{\partial^3M}{\partial S \partial Q_\alpha \partial Q_\beta}\Big)_{T=0}=
\fft{d}{dQ_\alpha}\Big[\Big(\fft{\partial^2 M}{\partial S \partial Q_\beta}\Big)_{T=0}\Big]-
 \Big(\fft{\partial^3M}{\partial S^2\partial Q_\beta}\Big)_{T=0} \fft{\partial S_\ext}{\partial Q_\alpha}\,.\label{WCCC3id}
\eea
The second and third identities have the origin of the Leibniz chain rule. Specifically, considering a function $F(S,Q)$, we have
\be
F(S,Q)\big|_{T=0}=F(S_\ext(Q),Q),\quad\rightarrow
\quad \fft{d}{dQ} \Big[F\big|_{T=0}\Big] =\Big(\fft{\partial F}{\partial Q}\Big)_{T=0} + \Big(\fft{\partial F}{\partial S}\Big)_{T=0} \fft{\partial S_{\ext}}{\partial Q} \,.
\ee
Note that the quantities in the square bracket in \eqref{WCCC3id} were given earlier in \eqref{WCCC2id}, indicating that there is an iterative process of expressing higher derivatives in terms of lower derivative results. Substituting \eqref{WCCC3id} into \eqref{WCCC30}, we have
\bea
X_\epsilon &\ge& \fft{1}{2W_1} (W_1 T_\e- \l \d S_\ext)^2 +
\fft1{6W_1} (W_1T_\e - \l \d S_\ext)\Big(
U_2W_1(W_1T_\e - \l \d S_\ext)^2 \nn\\
&&-3 \lambda(W_1 T_\e - \l \d S_\ext)\fft{\delta W_1}{W_1} + 3T_\epsilon^2 W_2 -3 \lambda^2 \delta^2 S_\ext\Big)+ {\cal O}(\lambda^4)\,.\label{WCCC31}
\eea
Since the quantity $(W_1 T_\epsilon - \lambda \delta S_\ext)$ can be both positive and negative, it appears that the right-hand side of the above inequality can be negative as $(W_1 T_\epsilon - \lambda \delta S_\ext)$ approaches zero either positively or negatively. However, as $X_\epsilon$ changes the sign, the fourth-order contribution cannot be ignored, which we shall discuss later.  For now, we simply require that the second-order term vanish, in which case, we have
\be
X_\epsilon \ge 0 + {\cal O}(\lambda^4)\,.
\ee

\subsubsection{The fourth and higher-order perturbations}

It is now straightforward to extend our discussion to all orders. The general differential identities that are used in the evaluation are iterative, given by
\bea
&&\Big(\fft{\partial^{k+1} M}{\partial S^{k+1}}\Big)_{T=0} = \Big(\fft{\partial ^k T}{\partial S^k}\Big)_{T=0} \equiv U_k(Q_\alpha)\,,\nn\\
&&\Big(\fft{\partial^{k+1} M}{\partial S^{k}\partial Q_\alpha}\Big)_{T=0} =
\fft{d}{dQ_\alpha} \Big[\Big(\fft{\partial^{k} M}{\partial S^{k}}\Big)_{T=0}\Big] - \Big(\fft{\partial^{k+1} M}{\partial S^{k+1}}\Big)_{T=0}\fft{\partial S_\ext}{\partial Q_\alpha}\,,\nn\\
&&\Big(\fft{\partial^{k+1} M}{\partial S^{k-1}\partial Q_\alpha\partial Q_\beta}\Big)_{T=0} =
\fft{d}{dQ_\alpha} \Big[\Big(\fft{\partial^{k} M}{\partial S^{k-1}\partial Q_\beta}\Big)_{T=0}\Big] - \Big(\fft{\partial^{k+1} M}{\partial S^{k}\partial Q_\beta}\Big)_{T=0}\fft{\partial S_\ext}{\partial Q_\alpha}\,,\label{gendiffid}\\
&&\qquad \vdots\nn
\eea
We find that when the lower bound of the second-order perturbation vanishes, namely
\be
\lambda \delta S_\ext = W_1 T_\epsilon\,,
\ee
the fourth-order perturbation yields
\begin{equation} \label{WCCC4}
	X_\e \geq \frac{1}{8 W_1} \left( W_2 T_\epsilon^2 - \lambda^2\d^2 S_\ext \right)^2 +\O(\lambda^5).
\end{equation}
Thus, the condition for WCCC at the fourth order is the same as that of the second order, depending only on the sign of $W_1=W$.

We performed the analysis up to and including the tenth order $(n = 10)$ and found similar results. To summarize, consider $n=2k$, when all the lower-bound of perturbations vanishes, namely
\be
\lambda^i \delta^i S_\ext = W_i T_\epsilon^i\,,\qquad i=1,2,\ldots, k\,.\label{WCCCcond4}
\ee
the next non-vanishing order perturbation is at the $2(k+1)^{\rm th}$ order, given by
\be
X_\epsilon \ge \fft{1}{2((k+1)!)^2 W_1}\left( W_{k+1} T_\epsilon^{k+1} - \lambda^{k+1}\d^{k+1} S_\ext \right)^2 + {\cal O}(\lambda^{2k+4})\,.\label{WCCCres}
\ee
Here all $\delta^i S_\ext$'s are defined {\it via} the expansion of the order parameter $\lambda$ in $S_\ext(\lambda) = S_\ext(Q(\lambda))$. Intriguingly, these results depend only on $W_i$, rather than $U_i$ defined in \eqref{gendiffid}. Although $U_i$ can be expressed in terms of $W_i$, we do not have to evaluate these identities since all the $U_i$'s simply drop out from the final expression \eqref{WCCCres}.

\subsection{Near saturation perturbations}

In the previous subsection, we arrived at \eqref{WCCCres} by assuming that the lower bounds associated with all the lower orders vanish, as in \eqref{WCCCcond4}. This is unnecessarily restrictive, and may leave loopholes in our derivation. We now consider the more general case, using the third-order perturbation \eqref{WCCC31} as an example. It has the form
\be
X_\epsilon \ge \fft{1}{W_1} \Big(\ft12(W_1 T_\e- \l \d S_\ext)^2 +
\ft1{6} (W_1T_\e - \l \d S_\ext) Y\Big) + {\cal O}(\lambda^4)\,.\label{WCCC32}
\ee
It is clear that if the quantity $(W_1 T_\e- \l \d S_\ext)^2$ is sufficiently large, the right-hand side above is positive. For $X_\epsilon$ to change the sign, $(W_1 T_\e- \l \d S_\ext)$ must be at order $\lambda^2$, since $Y$ is of order $\lambda^2$. Thus the possible sign change could only occur at
\be
W_1 T_\e- \l \d S_\ext = \lambda^2 c_2 + {\cal O}(\lambda^3)\,,
\ee
where $c_2$ is some order zero quantity. In this case, the non-vanishing order of $X_\epsilon$ is automatically fourth order, given by
\be
X_\epsilon \ge \fft{1}{8W_1} \Big(W_2 T_\epsilon^2 - \lambda^2 \delta^2 S_\ext + 2\lambda^2 c_2\Big)^2 + {\cal O}(\lambda^5)\,.
\ee
This has the same structure as \eqref{WCCC4}, where a more restrictive condition was imposed. The same situation holds at the fifth order, where the right-hand side takes the form
\be
X_\epsilon \ge \fft{1}{8W_1} \Big(W_2 T_\epsilon^2 - \lambda^2 \delta^2 S_\ext + 2\lambda^2 c_2\Big)^2 + \Big(W_2 T_\epsilon^2 - \lambda^2 \delta^2 S_\ext + 2\lambda^2 c_2\Big) \widetilde Y + {\cal O}(\lambda^6)\,,
\ee
where $\widetilde Y$ is of order $\lambda^3$. Thus for $X_\epsilon$ to change sign, the quantity $W_2 T_\epsilon^2 - \lambda^2 \delta^2 S_\ext + 2\lambda^2 c_2$ must be of order $\lambda^3$, in which case, the leading sixth order is again a total square with a coefficient of $1/W_1$. Our analysis indicates that this feature continues for all orders. Thus the condition \eqref{WCCCcond4} can be relaxed to be
\be
W_i T_\epsilon^i-\lambda^i \delta^i S_\ext = c_{i+1} \lambda^{i+1} + {\cal O}(\lambda^{i+2})\,,\qquad i=1,2,\ldots, k\,.\label{WCCCcond5}
\ee
where $c_{i+1}$'s are some order-zero quantities. The non-vanishing leading term of $X_\epsilon$ is then at the $2(k+1)^{\rm th}$ order, and it is always greater than a total square with a coefficient $1/W_1$.

\section{A direct one-step approach} \label{sec:direct}

In the previous section, we analyzed the WCCC in Gedankenexperiments order by order. Our analysis indicates that no matter how high an order we need to examine, it is always a total square. Furthermore, the higher-order terms also appear to contain the information of the lower-order ones when they become sufficiently small.  This strongly suggests that the full results, including all orders, may be subsumed into one total square. In this section, we show that this is indeed the case, in a simple one-step approach.

First, we note that $X_\epsilon$ defined in \eqref{Xepsilon} can be expressed as
\be\label{Xepsilon1}
X_\epsilon = M\big(S_\epsilon + \Delta S, Q + \Delta Q\big) -
M\big(S_\ext + \Delta S_\ext, Q + \Delta Q\big)\,.
\ee
Note that although $\Delta S$ includes perturbations of all orders, it is still a small quantity, and we have
\be
M\big(S_\epsilon + \Delta S, Q + \Delta Q\big) - M\big(S_\epsilon, Q + \Delta Q\big) = T(S_\epsilon, Q + \Delta Q) \Delta S + {\cal O}(\Delta S^2) \ge 0+ {\cal O}(\Delta S^2)\,.
\ee
Thus we have
\bea
X_\epsilon &\ge& M\big(S_\epsilon, Q + \Delta Q\big) -
M\big(S_\ext + \Delta S_\ext, Q + \Delta Q\big) + {\cal O}(\Delta S^2)\nn\\
&\ge& \fft1{2W_1} (S_\epsilon  -S_\ext - \Delta S_\ext)^2+
{\cal O}\Big((S_\epsilon -S_\ext - \Delta S_\ext)^3, \Delta S^2\Big)\,.
\eea
We thus have
\begin{equation} \label{X-1}
	X_\e \geq \frac{1}{2 W_1} \left[ \sum_{k=1}^{\infty} \frac{1}{k!} \left( W_k T_\e^k - \l^k \d^k S_\ext \right) \right]^2 + \cdots.
\end{equation}
This can reproduce all the results in the previous section. Further, it groups all the $\lambda$ and $T_\epsilon$ orders in one big square term; therefore, expansion orders of $\lambda$ and $T_\epsilon$ may not have to be homogeneous, but they can be independent.

\section{Kerr-Newman black hole revisited} \label{sec:KN}

We analyzed the WCCC Gedankenexperiments at all orders for a general class of black holes. In this section, we apply the results to the KN black hole and compare our results to those in the literature. In particular, WJ generalized the SW method to calculate arbitrary orders for the KN black hole \cite{Wang:2022umx}, but we find that our results disagree with \cite{Wang:2022umx} in some specific detail.

The outer horizon $ r_+ $ of the KN black hole is given by the larger root of the equation $r^2 + (J/M)^2 + Q^2 - 2M r = 0 $. The thermodynamic quantities of a KN black hole are expressed as
\begin{equation}
\p_a = \frac{J/M}{r_+^2 + (J/M)^2},\, \p_Q = \frac{r_+ Q}{r_+^2 + (J/M)^2}, \, T = \frac{r_+ - M}{2 \pi \left(r_+^2 + (J/M)^2 \right)},\, S = \pi \left(r_+^2 + (J/M)^2 \right).
\end{equation}
The extremal entropy $ S_\ext $ and the parameter $ W_1 $ are given by
\begin{equation}
	S_\ext = \pi \left( 2M_\ext^2 - Q^2 \right), \quad W_1 = \frac{4 \pi^2 (J^2 + M_\ext^4)}{M_\ext}.
\end{equation}
Using these relations, our result, as formulated in \eqref{WCCCres}, can be written as
\begin{equation} \label{we}
	X_\e \geq \frac{1}{2 ({k!})^2} \frac{M_\ext}{4 \pi^2 (J^2 + M_\ext^4)} \left[ W_k T_\epsilon^k -\lambda^k \d^k S_\ext \right]^2 +\O(\lambda^{2k+1})\,.
\end{equation}
WJ adopted a slightly different approach, generalizing the horizon condition as
\begin{equation} \label{h}
	0\le h(\l) \equiv \bar{M}(\l)^2 - \bar{M}(\l) \bar{Q}(\l) - \bar{J}(\l),
\end{equation}
where $ \bar{M} = M^2, \, \bar{Q} = Q^2 $ and $ \bar{J} = J^2 $. They derived a similar result, concluding that when the first $ (2k-2)^\text{th} $-order perturbations' lower bounds vanish, then the $(2k-1)^\text{th} $-order perturbation also vanishes, but the $ (2k)^\text{th} $-order perturbation becomes a perfect square, as
\begin{equation} \label{Jiang}
	h(\l) \geq \left( \frac{\l^k}{2 (k!)} \d^k Y \right)^2 +\O(\lambda^{2k+1})\,,
\end{equation}
where $ Y = 2 \bar{M}_\ext - \bar{Q} $, which equals to $ S_\ext/\pi $ in our notation. Therefore, compared to our \eqref{we}, we see that the $T_\epsilon^k$ contribution is missing in the above inequality \eqref{Jiang}. To be precise, there are two discrepancies between Eq.~\eqref{we} and Eq.~\eqref{Jiang}. The first concerns the overall coefficients, which arise from differences in the expansion functions and can be easily reconciled. The second involves the structure of the perfect square terms, where WJ made mistakes.

\subsection{Reconcilable coefficients}

To resolve the first discrepancy, we expand $ M(\l) $ near $ M_\ext(Q_i(\l)) $ in Eq.~\eqref{h}:
\begin{equation}
	h(\l) = h \left( M(\l), M_\ext(\l) \right) = h\left( M_\ext(\l), M_\ext(\l) \right) + \frac{\pd h}{\pd M} \bigg|_{M=M_\ext} \left( M(\l) - M_\ext(\l) \right) + \cdots,
\end{equation}
where $ h\left( M_\ext(\l), M_\ext(\l) \right) = 0 $ and $ M(\l) - M_\ext(\l) = X_\e $ as defined in Eq.~\eqref{Xepsilon}. Therefore,
\begin{equation}
	h(\l) = \frac{\pd h}{\pd M} \bigg|_{M=M_\ext} X_\e + \cdots, \qquad  \frac{\pd h}{\pd M} \bigg|_{M=M_\ext} = 2 M_\ext \left( M_\ext^2 + (J/M_\ext)^2 \right).
\end{equation}
With this transformation, the coefficients in Eq.~\eqref{we} and Eq.~\eqref{Jiang} are exactly matched.

\subsection{Structural discrepancy in the perfect squares}

Now we address the second inconsistency related to the perfect square terms. SW proposed the following inequality for the second-order correction \cite{Sorce:2017dst}
\begin{equation}
	\d^2 M - \p_\a \d^2 Q_\a \geq -\frac{\kappa}{8 \pi} \d^2 A,
\end{equation}
where $ \kappa $ is the surface gravity and $ A $ is the area of the event horizon. At the same time, the first-order correction is saturated, {\it i.e.}, $\d M - \p_\a \d Q_\a = 0 $. Therefore, this inequality is equivalent to our relations formulated in Eqs.~\eqref{deltaM}. WJ generalized SW's inequality to arbitrary order as
\begin{equation}
	\d^n M - \p_\a \d^n Q_\a \geq -\frac{\kappa}{8 \pi} \d^n A,
\end{equation}
which corresponds to our generalization in Eq.~\eqref{physcondn}. The discrepancy arises from WJ's use of an incorrect identity in their calculation, which led to an incorrect result for $ \frac{\kappa}{8 \pi} \d^n A $. Specifically, WJ used the following identity (Eq.~(B102) in \cite{Wang:2022umx})
\begin{equation}
	\At^2 - 2 \At M^2 + \At Q^2 + J^2 + \ft14{Q^2} = 0,
\end{equation}
where $ \At = \frac{A}{8 \pi}$. The correct identity should be
\begin{equation}
	\At^2 - 2 \At M^2 + \At Q^2 + J^2 + \ft14{Q^4} = 0.
\end{equation}
Using this, the correct result for $ \frac{\kappa}{8 \pi} \d^n A $ becomes
\begin{equation} \label{A}
	\begin{aligned}
		\frac{\kappa}{8 \pi} \d^n A &= \frac{1}{2M} \left[ \sum_{i=1}^{n-1} C^i_n \d^i M \d^{n-i} M - \frac{1}{2} \sum_{i=1}^{n-1} C^i_n \d^i Q \d^{n-i} Q - \frac{4 \pi}{A} \sum_{i=1}^{n-1} C^i_n \d^i J \d^{n-i} J \right. \\
		&\left. - \frac{\pi}{A} \left(\sum_{i=0}^{n} \sum_{j=0}^{n-i} \sum_{k=0}^{n-i-j} C_n^i C_{n-i}^j C_{n-i-j}^k \d^i Q \d^j Q \d^k Q \d^{n-i-j-k} Q - 4 Q^3 \d^n Q \right) \right].
	\end{aligned}
\end{equation}
This expression is equivalent to $ -\sum^{n-1}_{\a = 1} C^\a_{n-1} \d^\a \p_i \d^{n-\a} Q_i $ (Eqs.~\eqref{physcondn}), thereby verifying that the principle generalization used by WJ was consistent with our approach. Finally, with the corrected inequality relation, WJ's method will yield the same result as our Eq.~\eqref{we}.

\section{Conclusion}
\label{sec:con}

WCCC in Gedankenexperiments for a general class of black holes with zero-temperature extremal limit was analyzed up to and including the second order in \cite{Wu:2024ucf}. We follow the same strategy and extend the discussion to all orders. An analogous discussion in the literature is \cite{Wang:2022umx}, where higher-order perturbations for the KN black hole were analyzed. This provides us with a check of our general discussion. We find that the work in \cite{Wang:2022umx} is morally correct, but disagrees with our result in its detailed expression. After fixing an error in the mass/charge relation in \cite{Wang:2022umx}, we are able to redo the calculation and obtain the correct results that agree with our general formulae. We further devised a simpler one-step approach that led to a more general result that subsumes all the order-by-order results. What's intriguing is that the general result groups the quantities of all orders in one total square, as in \eqref{X-1}. This relaxes the requirement in the usual Gedankenexperiments that the two expansion orders, namely $T_\epsilon$ and $\lambda$, must be homogeneous, so the conclusion applies to the more general setting.

The upshot from our model-independent approach is that the WCCC in Gedankenexperiments for all orders is predicated on the positive sign of only one quantity, namely $W$ defined in \eqref{W1}, even though all the $W_k$ defined in \eqref{Wn} enter the testing inequality \eqref{X-1}. This conclusion does not only apply to the black holes in Einstein gravity, but also to all black holes in modified gravities with an extremal limit that satisfy the first law of black hole thermodynamics, since the only assumption made is the model-independent second law \eqref{secondlaw}. The physical meaning of this $W$ has been extensively discussed in \cite{Wu:2024ucf}. All the well-known black holes in the literature, most of which satisfy the no-hair theorem, have positive $W$ and hence they are safe in WCCC. Negative $W$ can only arise from black holes that violate the no-hair theorem, where WCCC might have a possibility of violation.  Our results, together with \cite{Wu:2024ucf}, turn the WCCC in Gedankenexperiments into a new intriguing question in black hole physics: What guarantees the positivity of $W$?

\section*{Acknowledgement}

This work was supported in part by NSFC (National Natural Science Foundation of China) Grants No.~12375052 and No.~11935009.

\appendix
\subsection*{Appendices}
\section{Notation and Conventions}
\label{app:notation}

In this appendix, we present the key notation, definitions, and the thermodynamic and perturbative framework employed throughout this paper for the analysis of WCCC using Gedankenexperiments.

\subsection{Thermodynamic Framework}

\begin{itemize}
	\item \textbf{Thermodynamic Ensemble and Quantities:} We work within the canonical ensemble where the black hole mass $M$ is considered a function of the entropy $S$ and a set of $N$ conserved charges $Q_\alpha$ ($\alpha = 1, \dots, N$), denoted as $M(S, Q_\alpha)$. Repeated Greek indices imply summation. Other standard thermodynamic quantities include temperature $T$, chemical potentials $\phi_\alpha = (\partial M / \partial Q_\alpha)$, horizon area $A$, and surface gravity $\kappa$.
	\item \textbf{Extremal Limit:} We consider black holes admitting a zero-temperature ($T=0$) extremal limit. In this limit, the mass and entropy become functions solely of the charges, denoted $M_{\text{ext}}(Q_\alpha)$ and $S_{\text{ext}}(Q_\alpha)$, satisfying $M_{\text{ext}}(Q_\alpha) = M(S_{\text{ext}}(Q_\alpha), Q_\alpha)$.
	\item \textbf{Near-Extremal State:} States close to extremality are characterized by a small temperature $T_\epsilon$. Thermodynamic quantities are expanded around their extremal values. The near-extremal entropy $S_\epsilon$ relates to $S_{\text{ext}}$ {\it via} an expansion in $T_\epsilon$:
	\begin{equation}
		S_\epsilon = S_{\text{ext}} + \sum_{k=1}^{\infty} \frac{1}{k!} W_k T_\epsilon^k,
	\end{equation}
	where $W_k$ are derivatives evaluated at $T=0$:
	\begin{equation}
		W_k \equiv \left( \frac{\partial^k S}{\partial T^k} \right)_{Q_\alpha; T=0}.
	\end{equation}
	Here, the differentiability of $S$ in the extremal limit is generally assumed. See Sec.~\ref{sec:orders} for further discussions.
	
	\item \textbf{Related Derivatives:} In order-by-order calculation, we also utilize derivatives of temperature with respect to entropy at $T=0$:
	\begin{equation}
		\Big(\fft{\partial^{k+1} M}{\partial S^{k+1}}\Big)_{T=0} = \Big(\fft{\partial^k T}{\partial S^k}\Big)_{T=0} \equiv U_k(Q_\alpha).
	\end{equation}
	Note that $U_1 = 1/W_1$. 
\end{itemize}

\subsection{Perturbation Scheme}

\begin{itemize}
	\item \textbf{Perturbation Parameter:} Perturbations induced by a test particle are tracked using an order parameter $\lambda$, which is used solely for bookkeeping. The near-extremal temperature $T_\epsilon$ is often treated as being of the same order as $\lambda$, {\it i.e.}~ $T_\epsilon \sim \lambda$.
	\item \textbf{Expansion:} A generic quantity $M$ undergoing perturbation is expanded as
	\begin{equation}
		M(\lambda) = M + \lambda \delta M + \frac{1}{2!} \lambda^2 \delta^2 M + \dots + \frac{1}{n!} \lambda^n \delta^n M + \cdots .
	\end{equation}
	The total perturbation is $\Delta M = M(\lambda) - M$.
	\item \textbf{Physical Process Condition:} The interaction is constrained by the second law of black hole thermodynamics, $\Delta S \geq 0$. In the order-by-order analysis, this translates to: the $n$-th order condition $\delta^n S \geq 0$ is imposed when the conditions for all lower orders are saturated ($\delta S = \dots = \delta^{n-1} S = 0$). Applying this condition to the perturbative expansion of the mass function \eqref{MSQ} yields
	\bea
	&&\delta^i M - \phi_\alpha \delta^i Q_\alpha = Y^{\sst{(i)}}\,,\qquad i=1,2,\ldots, n-1\,,\nn\\
	&&\qquad\qquad\delta^n M - \phi_\alpha \delta^n Q_\alpha \ge  Y^{\sst{(n)}}\,.
	\eea
	where
	\be
	Y^\1=0\,,\qquad Y^{\sst{(k)}} =\sum^{k-1}_{i = 1} C^i_{k-1} \d^i \phi_\alpha \d^{k-i} Q_\alpha\,,\qquad k\ge 2\,.
	\ee
	Here, $C^i_j $ with $j\ge i$ is the binomial coefficient.
\end{itemize}

\subsection{WCCC Test}

\begin{itemize}
	\item \textbf{Horizon Condition Quantity:} The potential violation of WCCC is assessed via the sign of $X_\epsilon$, defined as
	\begin{equation}
		X_\epsilon \equiv M(S_\epsilon, Q_\alpha) + \Delta M - M_{\text{ext}}(Q_\alpha + \Delta Q_\alpha).
	\end{equation}
	The conjecture is upheld if the final state corresponds to a black hole or avoids a naked singularity, which requires $X_\epsilon \geq 0$.
	\item \textbf{General Result Structure:} The analysis across all orders, particularly the one-step approach in Sec.~\ref{sec:direct} culminating in \eqref{X-1}, reveals that the derived lower bound on $X_\e$ takes the form:
	\begin{equation}
		X_\e \geq \frac{1}{2 W_1} \left[ \sum_{k=1}^{\infty} \frac{1}{k!} \left( W_k T_\e^k - \l^k \d^k S_\ext \right) \right]^2 + \cdots.
	\end{equation}
	This structure demonstrates that $W_1 > 0$ is sufficient to ensure $X_\epsilon \geq 0$ under the examined perturbations.
\end{itemize}

\bibliographystyle{JHEP}

\end{document}